\documentclass[intlimits,twoside,a4paper]{article}

\usepackage[cp1251]{inputenc}

\usepackage{cmpj3}

\issue{2018}{21}{1}{13703}
\doinumber{10.5488/CMP.21.13703}

\title[Optical properties of GaAs/Al$_{x}$Ga$_{1-x}$As/GaAs]%
{Optical properties of GaAs/Al$_{x}$Ga$_{1-x}$As/GaAs quantum dot with off-central impurity driven by electric field}
\author[V.A.~Holovatsky, M.Ya.~Yakhnevych, O.M.~Voitsekhivska]{V.A.~Holovatsky\footnote{E-mail: ktf@chnu.edu.ua}\,,
M.Ya.~Yakhnevych, O.M.~Voitsekhivska}
\address{Yuriy Fedkovych Chernivtsi National University, 2 Kotsyubynsky St.,
58012 Chernivtsi, Ukraine}

\date{Received November 24, 2017, in final form December 26, 2017}

\begin{document}
\maketitle

\begin{abstract}
The effect of a constant electric field and donor impurity on the energies and oscillator strengths of electron intraband quantum transitions in double-well spherical quantum dot GaAs/Al$_{x}$Ga$_{1-x}$As/GaAs is researched. The problem is solved in the framework of the effective mass approximation and rectangular potential wells and barriers model using the method of wave function expansion over a complete set of electron wave functions in nanostructure without electric field. It is shown that under the effect of electric field, the electron in the ground state tunnels from the inner potential well into the outer one. It also influences on the oscillator strengths of intraband quantum transition. The binding energy of an electron with ion impurity is obtained as a function of electric field intensity at a different location of impurity.


\pacs 71.38.-k, 63.20.kd, 63.20.dk, 72.10.Di
\end{abstract}

\section{Introduction}

Multishell quantum dots (QDs) are new and very interesting objects of research in nanostructure physics. The existence of several potential wells with sizes, which can be changed in the processes of nanostructure growth, allows one to create multimodal sources and detectors of electromagnetic waves. The relationships between the sizes of the potential wells determine the location of quasi-particles. The external fields significantly change the energy spectra and location of quasi-particles too. For example, in  \cite{Hol16, Hol17, Holo17, Hol14} it is shown that the electron location is changed under the effect of a magnetic field. The electron in the ground state tunnels from the outer potential well into the inner one. This process is accompanied by the varying oscillator strengths of intraband transitions.

The study of the impurity states in semiconductor nanostructures was initiated only in early 1980s through the pioneering work of Bastard \cite{Bas81}. In spite of the growing interest to the topic of impurity doping in nanocrystallites, the majority of theoretical works have been carried out on shallow donors in spherical QDs employing perturbation methods \cite{Bos99, Bos98, Bose98, BoseC98, Yua08} or variational approaches \cite{Zhu90, Zhu94, Joh05, Kan06, Man05, Dan08}. For example, using the perturbation methods, Bose et al. \cite{Bos99, Bos98, Bose98, BoseC98} obtained the binding energy of a shallow hydrogenic impurity in spherical QDs. Based on variational approaches, Zhu \cite{Zhu90, Zhu94} studied the energies of an off-center hydrogenic donor confined by a spherical QD with a finite rectangular potential well. Using the plane wave method, Li \cite{Lis07} calculated the electronic states of hydrogenic donor impurity in low-dimensional semiconductor nanostructures in the framework of the effective mass envelope-function theory. In \cite{Kos12}, the influence of central impurity and external electric field on the energies of intraband quantum transitions and absorption coefficient is studied in spherical quantum dot CdTe/ZnTe.

The effect of impurity on the energy spectra in multishell nanostructures is researched within different methods in \cite{Cri12, Hol09, Holo14, Hol13, Boi12, Boi10, Boi11}. It is shown that electron-impurity binding energy depends on the sizes of nanostructure and impurity location. The external electric field changes the most probable electron location and increases or decreases the binding energy depending on the impurity location. Besides, in multishell nanostructures, the electron changes its location under the effect of an electric field. That is manifested by the optical properties and can be used in new optoelectronic devices. The effect of electric field on quasi-particles energy spectra in spherical nanostructures with one potential well is investigated in \cite{Holov14, Nic13, Kos12}.

In this paper we study the electron energy spectrum and the wave function of its ground state as functions of the intensity of electric field taking into account the polarization effects, which arise due to the presence of interfaces between nanostructure materials. The oscillator strengths of intraband quantum transitions and binding energy of electron with donor impurity are calculated at its different location.

\section{Theoretical model}

A semiconductor spherical QD consisting of a core-well with the radius $r_{0} $, a barrier of the thickness $\Delta =r_{1} -r_{0} $ and a well of the width $\rho \, =r_{2} -r_{1} $ placed into the semiconductor matrix-barrier is studied. In order to investigate the effect of the electric field on the electron energy spectrum and wave functions in the nanostructure with impurity, the Schr\"odinger equation with the Hamiltonian
\begin{equation} \label{GrindEQ__1_}
H=-\vec{\nabla }\frac{\hbar ^{2} }{2\mu(r)}\vec{\nabla}+V_{F} (r,\theta) +V_{\text c} (\vec{r}) +V_{\text p}(r,\theta)+W\left(r\right)+U(r)
\end{equation}
is solved.
The confining potential $U(r)$ and the effective mass $\mu (r)$ are step-like functions:
\begin{equation} \label{GrindEQ__5_}
U(r)=\left\{\begin{array}{l} {0, \, \qquad r\leqslant r_{0}\,,\  r_{1} <r\leqslant r_{2}\,, } \\ {V,  \qquad r_{0} <r\leqslant r_{1}\,, } \\ {\infty, \qquad r>r_{2}\,, } \end{array}\right.
\end{equation}
\begin{equation} \label{GrindEQ__6_}
\mu (r)=\left\{\begin{array}{l} {m{}_{0}\,, \, \qquad r\leqslant r_{0}\,,\ \ r_{1} <r\leqslant r_{2}\,, \qquad \text{wells},} \\ {m_{1}\,, \, \qquad r_{0} <r\leqslant r_{1}\,,  \qquad \qquad \quad \text{barrier}.} \end{array}\right. 
\end{equation}
 The potential energy of interaction between the electron and the positive ion, which is located at $z$-axis at the distance $r_{\text{imp}}$ from the center of the nanostructure, has the form
\begin{equation} \label{GrindEQ__3_}
V_{\text c} (\vec{r})=-\frac{Z e^{2} }{\varepsilon |\vec{r}-\vec{r}_{\text{imp}} |}\,,
\end{equation}
where $\varepsilon =\sqrt{\varepsilon _{1} \, \varepsilon _{2} } $ is an average dielectric constant, $\varepsilon _{1} $, $\varepsilon _{2} $ are the dielectric constants of semiconductor materials of the wells and barriers, respectively.

 \begin{equation} \label{GrindEQ__3a_}
V_{\text p} \left({r,\theta}\right)=-\frac{Z e^{2}(\varepsilon-\varepsilon _{3})}{\varepsilon r_{2}}\sum_{k=0}^{\infty}\frac{r_{i}^{k}r^{k}}{r_{2}^{2k}} \frac {k+1}{k\varepsilon + (k+1)\varepsilon_{3}} P_{k}(\cos\theta) .
\end{equation}
 Here, $\varepsilon _{3} $ is the dielectric constant of semiconductor matrix-barrier and $P_{k}(\cos\theta)$  is the Legendre polynomial. A formula arises from the existence of the polarized surface charges at the dot boundary and describes the interaction between the ion and electron \cite{Nicu13}.

$W(r)$ term in formula~\eqref{GrindEQ__1_} describes the electron self-polarization potential, which in the case of a small difference between $\varepsilon _{1} $ and $\varepsilon _{2} $ is simplified to the following form \cite{ Nicu13, Nic13}
\begin{equation} \label{GrindEQ__4_}
W(r)=\frac{e^{2} (\varepsilon -\varepsilon _{3} )}{8\piup \varepsilon  r_{2} } \sum _{k=0}^{\infty }\frac{k+1}{k \varepsilon +(k+1)\varepsilon _{3} } (r/r_{2} )^{2k}.
\end{equation}

 Here, $V_{F}(\vec{r})$ is an electrostatic potential of the electron in external field $\vec{F}$ applied in $z$-direction
\begin{equation} \label{GrindEQ__2_}
V_{F} (r,\theta )=-e F\cos \theta \left\{\begin{array}{l} {a_{0} r,\qquad \qquad\, \ r\leqslant r_{0}\,, } \\ {a_{1} r+ \displaystyle \frac{b_{1} }{r^{2}}\,,\qquad r_{0} <r\leqslant r_{1} \,, } \\[6pt] {a_{2} r+ \displaystyle \frac{b_{2} }{r^{2} }\,, \qquad\, r_{1} <r\leqslant r_{2} \,, } \\ {r+ \displaystyle \frac{b_{3} }{r^{2} }\,, \qquad \quad\, r>r_{2} . } \end{array}\right.
\end{equation}

 The coefficients $a_{i}$ and $b_{i}$ (presented in the appendix~\ref{app}) are obtained as solutions of Poisson equation with standard dielectric boundary conditions at the interfaces.

 If $F=0$, $Z=0$, the Schr\"{o}dinger equation with Hamiltonian~\eqref{GrindEQ__1_} has exact solutions
\begin{equation} \label{GrindEQ__7_}
\Phi _{n\ell m} (r,\theta ,\varphi )=R_{n\ell} (r)Y_{\ell m} (\theta ,\varphi ),
\end{equation}
where $Y_{\ell m} (\theta ,\varphi )$ are spherical functions, and the radial wave functions are written as:
\begin{equation} \label{GrindEQ__8_}
R_{n\ell}^{(i)}(r) =A_{n\ell}^{(i)} J_{\ell}^{(i)} (\kappa _{n\ell} r)+B_{n\ell}^{(i)} N_{\ell}^{(i)} (\kappa _{n\ell} r),  \qquad i=0,1,2,
\end{equation}
\begin{equation} \label{GrindEQ__9_}
J_{\ell}^{(i)} (\chi _{n\ell} r)=\left\{\begin{array}{l} {j_{\ell} (k_{n\ell} r), \ \quad  i=0,2,} \\ {I_{\ell} (\chi _{n\ell} r), \ \quad i=1,}\end{array}\right.
\end{equation}
\begin{equation} \label{GrindEQ__10_}
N_{\ell}^{(i)} (\chi _{n\ell} r)=\left\{\begin{array}{l} {n_{\ell} (k_{n\ell} r), \ \quad  i=0,2,} \\ {K_{\ell} (\chi _{n\ell} r), \quad i=1}. \end{array}\right.
\end{equation}
Here, $j_{\ell }$, $n_{\ell } $ are spherical Bessel functions of the first and the second kind, $I_{\ell }$, $K_{\ell } $ are modified  spherical Bessel functions of the first and the second kind, $k_{n\ell } =({2m_{0} E_{n\ell }^{0} /\hbar })^{1/2}  $, $\chi _{n\ell } =[{2m_{1} (V-E_{n\ell }^{0} )/\hbar }]^{1/2}  $.

 The unknown coefficients $A_{n\ell }^{(i)}$, $B_{n\ell }^{(i)} $ and the electron energies $E_{n\ell }^{0} $ in QD are obtained using the continuity conditions for the wave functions and their densities of currents at all interfaces:
\begin{equation} \label{GrindEQ__11_}
\left. \begin{array}{l} {R_{n\ell }^{(i)} \left(r_{i} \right)=R_{n\ell }^{(i+1)} \left(r_{i} \right)} \\[6pt] {\left. \displaystyle \frac{1}{m_{i} } \displaystyle\frac{\rd R_{n\ell }^{(i)} \left(r\right)}{\rd r} \right|_{r=r_{i} } =\left. \displaystyle \frac{1}{m_{i+1} } \displaystyle \frac{\rd R_{n\ell }^{(i+1)} \left(r\right)}{\rd r} \right|_{r=r_{i} } } \end{array}\right\} \quad i=0,1,2
\end{equation}
and the normalization condition for the radial wave function
\begin{equation} \label{GrindEQ__12_}
\int _{0}^{\infty }\left|R_{n\ell } (r)\right|^{2} r^{2} \rd r=1.
 \end{equation}

The coefficients $B_{n\ell }^{(0)} =0$,  $A_{n\ell }^{(3)} =0$ from the condition that the wave function is finite at $r=0$ and $r\to \infty $,  respectively.
In order to study the electron properties in a nanostructure driven by the electric field, we are going to use the method of expansion of the quasi-particle wave function using a complete set of eigenfunctions of the electron in a spherical nanostructure without the external fields obtained as the exact solutions of Schr\"{o}dinger equation. When the external fields are applied, the spherical symmetry of the problem is broken and the orbital quantum number $\ell$ becomes not a valid one. Now, the quantum number \textit{j} determines the number of the energy level with a fixed magnetic quantum number \textit{m}.  The new states characterized by a magnetic quantum number \textit{m} are presented as a linear combination of the states characterized by the functions $\Phi _{n\ell m} (\vec{r})$
\begin{equation} \label{GrindEQ__13_}
\psi _{jm} (\vec{r})=\sum _{n}\sum _{\ell}C_{n \ell} \Phi _{n \ell m} (\vec{r }).
\end{equation}
Substituting~\eqref{GrindEQ__8_} into the Schr\"{o}dinger equation with Hamiltonian~\eqref{GrindEQ__1_}, we obtain a secular equation
\begin{equation} \label{GrindEQ__14_}
\left|H_{n \ell,n'\ell'} -E_{jm} \delta _{n,n'} \delta _{\ell,\ell'} \right|=0.
\end{equation}
In order to obtain the electron energy spectrum and wave functions under the effect of external fields and impurity, one should calculate the eigenvalues and eigenvectors of the matrix. Herein,
\begin{equation} \label{GrindEQ__15_}
H_{n \ell ,n'\ell '} =E_{n\ell }^{0} \delta _{n',n} +\xi (\alpha _{\ell ,m} \delta _{\ell ',\ell +1} +\beta _{\ell ,m} \delta _{\ell ',\ell -1} ) U_{n'\ell ',n\ell } +I _{n'\ell ',n\ell } \delta _{\ell ',\ell }\, ,
\end{equation}
\[I_{n'\ell ',n\ell } =\int _{V}\Phi _{n'\ell '}^{*} (\vec{r })V_{\text c} (\vec{r })\Phi _{n\ell } (\vec{r })\rd\vec{r },\qquad U_{n'\ell ',n\ell } =\int _{0}^{\infty }r^{2} R_{n'\ell '}^{*} (r)R_{n\ell } (r)\rd r, \]
\[\alpha _{\ell ,m} =\sqrt{\frac{\ell ^{2} -(m+1)^{2} }{(2\ell +3)(2\ell +1)} }\,, \qquad \beta _{\ell ,m} =\sqrt{\frac{\ell ^{2} -m^{2} }{4\ell ^{2} -1} }\, ,\]
where $\xi =eF\varepsilon$ and $\vec{r}=(r,\theta ,\varphi )$.

 In the nanostructure driven by an electric field, the intensity of intraband quantum transitions is given by the formula
\begin{equation} \label{GrindEQ__16_}
I_{i{-}f} \sim (E_{i} -E_{f} )\Bigg|\int _{V}\psi _{i}^{*} (\vec{r}) r\cos \theta  \psi _{f} (\vec{r}) \rd V\Bigg|^{2} .
\end{equation}
In this paper, the energies and the intensities of quantum transition are found as functions of the electric field strength.

The probability of intraband quantum transitions is determined by the oscillator strength \[F_{i-f}=2m\omega_{if}|d_{if}|^{2}/\hbar e^{2}\] ($\omega _{if}$ is the frequency and $d_{if} $ is the dipole momentum of the transition). Using the atomic energy units, the radial coordinate and taking into account the coordinate-dependent electron effective mass, the formula for oscillator strength of the transitions from the ground state ($j=1$, $m=0$) takes the form \cite{Hol17}:
\begin{equation} \label{GrindEQ__18_}
F_{i-f} =(E_{i} -E_{f} )\left|{\left\langle i \right|} \sqrt{\mu (r)} \, r\cos \theta  {\left| f \right\rangle} \right|^{2}.
\end{equation}

The Thomas-Reiche-Kuhn sum rule must be fulfilled
\begin{equation} \label{GrindEQ__19_}
\sum _{f}F_{i-f}  =1,
\end{equation}
where $i$ denotes the initial state  and $f$ is the final state of the transition. It gives an opportunity to find all quantum transitions and controls the accuracy of numerical calculations.

\section{Results and discussion}

The computer calculations were performed using the physical parameters of Al$_{x}$Ga$_{1-x}$As semiconductor with Al concentration $x=0$ for the potential wells and $x=0.4$ for the barriers, $m_{0} =0.067 m_{e} $, $m_{1} =0.1 m_{e} $, $V=297$~meV, $\varepsilon _{1} =13.2$, $\varepsilon _{2} =11.9$, $\varepsilon _{3} =1$, where $m_{e} $ is the mass of a pure electron. The sizes of potential wells are selected in such a way that the electron should be located in the core of a nanostructure without an electric field and impurity ($r_{0}=6$~nm, $\rho=5$~nm, $\Delta=2$~nm).

\begin{figure}[!t]
\centerline{\includegraphics[width=0.7\textwidth]{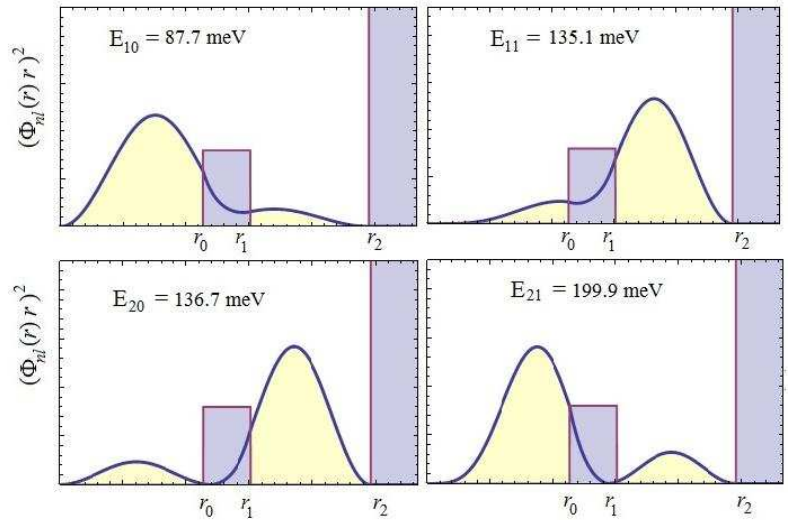}}
\caption{\label{fig1} (Colour online) The radial distribution of probability density for the electron in the states $|10\rangle$, $|11\rangle$, $|20\rangle$ and $|21\rangle$.}
\end{figure}

In figure~\ref{fig1}, the radial distributions of probability densities for the electron in the ground and three excited states are presented. We took into account not less than 36 terms in the expansion~\eqref{GrindEQ__8_} ($n=1,\ldots,6$ and $\ell=0,\ldots,5$). When the electric field intensity increases and the impurity shifts from QD center, the number of states $|n \ell m\rangle$, which make a significant contribution into $\psi _{jm} (\vec{r})$, increases.

\begin{figure}[!t]
\centerline{\includegraphics[width=0.68\textwidth]{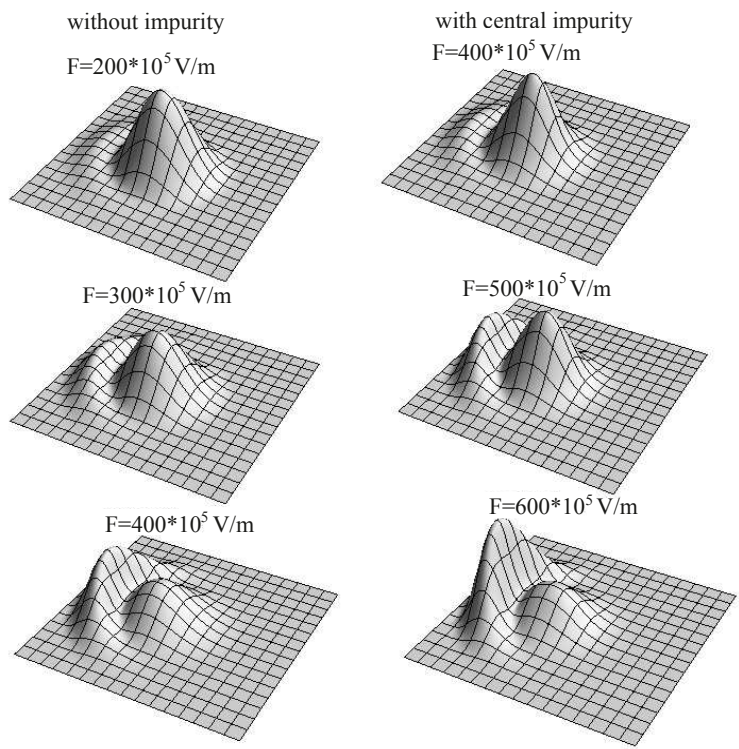}}
\vspace{-3mm}
\caption{\label{fig2} Distribution of electron density in the ground state in a nanostructure with and without impurity at different values of the electric field intensity ($F$).}
\end{figure}

 The results of $\left|\psi _{10} (\vec{r})\right|^{2} $calculations show, figure~\ref{fig2}, that the electron, at certain geometric parameters, is located in the core of a nanostructure. When the electron is driven by an electric field, it tunnels through the barrier into the outer potential well. Moreover, when a central impurity is present, this transition happens at a bigger value of the electric field intensity, which compensates the Coulomb attraction between the impurity and electron.

\begin{figure}[!t]
\centerline{\includegraphics[width=0.52\textwidth]{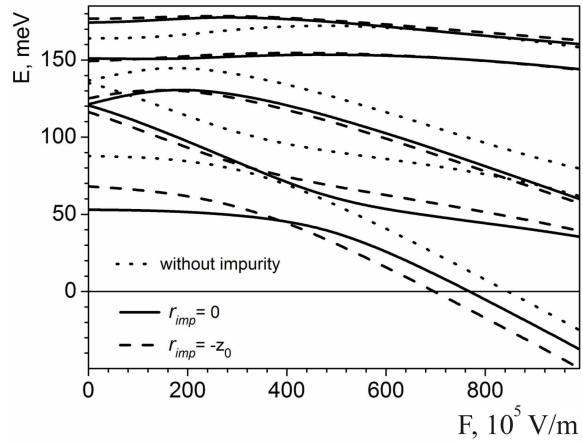}}
\caption{\label{fig3} Electron energies as functions of the electric field intensity in QD with on-center impurity (solid curves), at $r_{\text{imp}}=-z_{0}$ (dashed curves) and without impurity (dotted curves).}
\end{figure}

 \begin{figure}[!t]
\centerline{\includegraphics[width=0.52\textwidth]{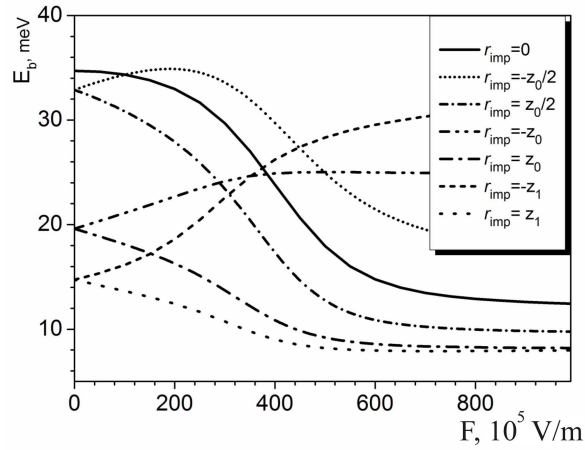}}
\caption{\label{fig4} Electron-impurity binding energy as a function of the electric field intensity at $r_{\text{imp}}=-z_{1}$, $-z_{0}$, $-z_{0}/2$, 0, $z_{0}/2$, $z_{0}$, $z_{1}$.}
\end{figure}

The electron energies as functions of the electric field intensity ($F$) are presented in figure~\ref{fig3}. It is clear that the change of the electron location is accompanied by the effect of anti-crossing energy levels. If the impurity is located at the distance $r_{\text{imp}}=-z_{0}$, the anti-crossing of energies happens at the same intensity of the electric field the same as without impurity. The ground state energy of an electron, located in the outer potential well, linearly decreases at a bigger intensity.

Under the influence of the electric field, the electron changes its most probable location generating, in its turn, a varying binding energy with the ion impurity which is determined by the formula
\begin{equation} 
E_{\text b} =E_{10}-E_{10}^{\text{imp}} ,
\end{equation}
where $E_{10} $  is the energy of electron ground state without impurity, $E_{10}^{\text{imp}} $ --- that with the impurity.

In figure~\ref{fig4}, the electron-impurity binding energy as a function of the electric field intensity at $r_{\text{imp}}=-z_{1}$, $-z_{0}$, $-z_{0}/2$, 0, $z_{0}/2$, $z_{0}$, $z_{1}$ is shown. The figure proves that at $z_{0} \geqslant 0$, the binding energy monotonously decreases because the electron moves from the impurity, but at $z_{0} <0$, the function has a non-monotonous character, because the electron at first moves to the impurity and then from it.

\begin{figure}[!t]
\centerline{\includegraphics[width=0.58\textwidth]{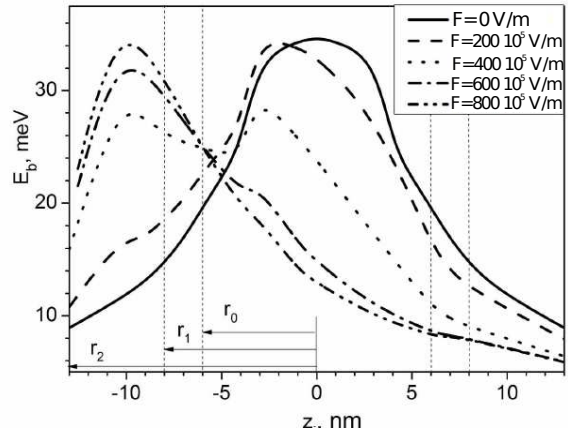}}
\caption{\label{fig5} Electron-impurity binding energy as a function of the electron location.}
\end{figure}

\begin{figure}[!t]
\centerline{\includegraphics[width=0.52\textwidth]{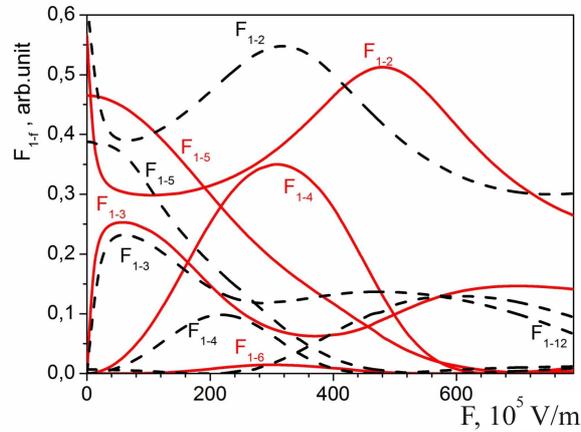}}
\caption{\label{fig6} (Colour online) Oscillator strengths of intraband quantum transitions as functions of the electric field intensity. Solid curves --- with impurity, dashed curves --- without impurity.}
\end{figure}

In figure~\ref{fig5}, the dependences of the binding energy of an electron with impurity on its location are shown at different values of the electric field intensity. An increasing electric field intensity shifts the maximum of the binding energy and weakly changes its magnitude.

The results of calculations of oscillator strengths of intraband quantum transitions are presented in figure~\ref{fig6}. These non-monotonous dependences are explained by a varying symmetry of wave functions, and their overlapping due to the electron changes its location.

In figure~\ref{fig6}, the oscillator strengths of intraband quantum transitions are presented as functions of the electric field intensity. The numeric calculations are performed at a magnetic quantum number $m=0$, thus we denote the states only by a quantum number $j=1,2,3,\ldots .$ From the figure one can see that the quantum transition between the ground and the first excited state ($|1\rangle\to|2\rangle$) is the most intensive in the whole range of the electric field intensity when there is no impurity. When the central impurity appears, besides the transition $|1\rangle\to|2\rangle$, the oscillator strengths of quantum transitions into the states with bigger energies ($|1\rangle\to|3\rangle$, $|1\rangle\to|4\rangle$, $|1\rangle\to|5\rangle$) essentially increase. These optical properties of multishell nanostructures can be used in semiconductor devices.

\section{Summary}

The effect of an electric field and a shallow charged impurity on the intraband quantum transition of an electron in the spherical multishell QD with two potential wells is investigated. The Schr\"{o}dinger equation is solved using the matrix method and the effective mass approximation, taking into account the polarization effects. We studied how the electron changes its location in a nanostructure driven by an electric field. The sizes of potential wells are selected such that the electron, in the ground state, is to be located in the core of nanostructure without electric field and impurity. Under the effect of an electric field the depth of the outer potential wells increases, because the electron tunnels from the core-well into the outer one. The electron changes its location which manifests in the complicated behavior of oscillator strengths of intraband quantum transitions depending on the electric field intensity.

\appendix
\section{Solutions of Poisson equation} \label{app}
\vspace{-6mm}
\begin{align}
a_{0} &=27r_{1}^{3} r_{2}^{3} \varepsilon _{1}\varepsilon _{2} \varepsilon _{3} \big\{2r_{0}^{3} (\varepsilon _{1} -\varepsilon _{2} )\big[r_{1}^{3} (2\varepsilon _{1} +\varepsilon _{2} )(\varepsilon _{1} -\varepsilon _{3} )+r_{2}^{3} (\varepsilon _{2} -\varepsilon _{1} )(2\varepsilon _{3} +\varepsilon _{1})\big]
\nonumber\\
&+r_{1}^{3} (2\varepsilon _{1} +\varepsilon _{2} )\big[2r_{1}^{3} (\varepsilon _{2} -\varepsilon _{1} )(\varepsilon _{1} -\varepsilon _{3} )+r_{2}^{3} (\varepsilon _{2} +2\varepsilon _{1} )(2\varepsilon _{3} +\varepsilon _{1} )\big]\big\}^{-1}\nonumber,\\
a_{1} &=\frac{a_{0} (2\varepsilon _{2} +\varepsilon _{1} )}{3\varepsilon _{2} }\,, \qquad b_{1} =\frac{r_{0}^{3} (\varepsilon _{2} -\varepsilon _{1} )}{\varepsilon _{1} +2\varepsilon _{2} }\,, \qquad a_{2} =\frac{a_{1} \left[2b_{1} (\varepsilon _{1} -\varepsilon _{2} )+r_{1}^{3} (\varepsilon _{2} +2\varepsilon _{1} )\right]}{3r_{1}^{3} \varepsilon _{1} }\,, \nonumber
\end{align} 
\begin{equation}
b_{2} =\frac{r_{1}^{3} \left[r_{1}^{3} (\varepsilon _{1} -\varepsilon _{2} )+b_{1} (\varepsilon _{1} +2\varepsilon _{2} )\right]}{2b_{1} (\varepsilon _{1} -\varepsilon _{2} )+r_{1}^{3} (2\varepsilon _{1} +\varepsilon _{2} )}\,, \qquad b_{3} =\frac{2a_{2} b_{2} \varepsilon _{1} -a_{2} r_{2}^{3} \varepsilon _{2} +r_{2}^{3} \varepsilon _{3} }{2\varepsilon _{3} }.\nonumber
\end{equation}

If $\varepsilon_{1}\approx\varepsilon_{2}$ and $\varepsilon =\sqrt{\varepsilon _{1} \, \varepsilon _{2} } $, the coefficients $a_{i}$ and $b_{i}$ have the form
\begin{equation}
a_{0} =a_{1}=a_{2}=\frac{3 \varepsilon_{3}}{\varepsilon +\varepsilon_{3}}\,,\qquad b_{1}= b_{2} =0, \qquad b_{3} =\frac{r_{2}^{3}( \varepsilon_{3}-\varepsilon)}{\varepsilon +2\varepsilon_{3}}. \nonumber
\end{equation}

\newpage
\ukrainianpart

\title{Оптичні властивості квантової точки GaAs/Al$_{x}$Ga$_{1-x}$As/GaAs з нецентральною домішкою під впливом електричного поля }
\author{В.А. Головацький, М.Я. Яхневич, О.М. Войцехівська}
\address{Чернівецький національний університет ім. Ю.~Федьковича,\\ вул. Коцюбинського, 2,  58012 Чернівці, Україна }

\makeukrtitle

\begin{abstract}
Досліджено вплив постійного електричного поля та донорної домішки на енергії та сили осцилятора внутрішньозонних квантових переходів електрона в двоямній сферичній квантовій точці GaAs/Al$_{x}$Ga$_{1-x}$As/GaAs. Задача розв'язана в рамках наближення ефективних мас та моделі прямокутних потенціальних  ям і бар'єрів методом розкладу хвильової функції за повним набором хвильових функцій електрона у наносистемі без електричного поля. Показано, що під   впливом електричного поля електрон в основному стані може тунелювати з внутрішньої потенціальної ями в зовнішню. Це супроводжується зміною сил осциляторів внутрішньозонних квантових переходів. Отримано залежність енергії зв'язку електрона іоном домішки від напруженості електричного поля при різних
положеннях домішки.
\keywords  багатошарова квантова точка, домішка, внутрішньозонні квантові переходи

\end{abstract}

\end{document}